\documentclass[aps,prb,twocolumn,floatfix,superscriptaddress,showpacs,amsmath,amssymb]{revtex4}
\usepackage{graphicx}
\usepackage{epsfig}  
\usepackage{xcolor}
\allowdisplaybreaks

\newcommand{\be}{\begin{equation}}
\newcommand{\ee}{\end{equation}}
\newcommand{\bea}{\begin{eqnarray}}
\newcommand{\eea}{\end{eqnarray}}
\newcommand{\ba}{\begin{eqnarray*}}
\newcommand{\ea}{\end{eqnarray*}}

\newcommand{\bR}{\mathbf{R}}

\newcommand{\bq}{\mathbf{q}}

\newcommand{\bRp}{\mathbf{R'}}

\newcommand{\m}[1]{\mathcal{#1}} 

\newcommand{\spin}[1]{\sigma^{#1}}

\newcommand{\omo}{\omega_{0}} 	% Resonant case \omr=\omq=\omo

\newcommand{\Fig}[1]{Fig.~\ref{#1}}

\begin{document}
 
\title{Quantum Phase Transition of Light in the Rabi-Hubbard Model}
\author{M. Schir\'o}
\affiliation{Princeton
 Center for Theoretical Science and Department of Physics, Joseph Henry
 Laboratories, Princeton University, Princeton, NJ
  08544, USA}
\author{M. Bordyuh}
\affiliation{Department of Electrical Engineering, Princeton University, Princeton, NJ
  08544, USA}
 \author{B. \"{O}ztop}
 \affiliation{Institute for Quantum Electronics, ETH-Z\"urich, CH-8093 Z\"urich,
Switzerland}
\affiliation{Department of Electrical Engineering, Princeton University, Princeton, NJ
  08544, USA}
  \author{H. E. T\"ureci}
\affiliation{Department of Electrical Engineering, Princeton University, Princeton, NJ
  08544, USA}

\date{\today} 
 
\begin{abstract}
We discuss the physics of the Rabi-Hubbard model describing large arrays of coupled cavities interacting with two level atoms via a Rabi non-linearity. We show that the inclusion of counter-rotating terms in the light-matter interaction, often neglected in theoretical descriptions based on Jaynes-Cumming models, is crucial to stabilize finite-density quantum phases of correlated photons with no need for an artificially engineered chemical potential.  We show that the physical properties of these phases and the quantum phase transition occurring between them is remarkably different from those of interacting bosonic massive quantum particles. The competition between photon delocalization and Rabi non-linearity drives the system across a novel  $Z_2$ parity symmetry-breaking quantum phase transition between two gapped phases, a Rabi insulator and a delocalized super-radiant phase. 
\end{abstract}

\maketitle
\section{Introduction}

Recent years have seen an enormous increase in the level of control over light-matter interactions at the quantum level with both atomic and solid-state systems~\cite{haroche_exploring_2006, yamamoto_semiconductor_2000, schoelkopf_wiring_2008}. These major experimental achievements have brought forth a novel class of systems where light and matter play equally important roles in emergent quantum many body phenomena.  The basic building block of such systems is the elementary Cavity QED (CQED) system formed by a two-level system (TLS) interacting with a single mode of an electromagnetic resonator. 

When CQED systems are coupled to form a lattice a plethora of novel phenomena are expected to emerge. The possibility of  quantum phase transitions of light between Mott-like insulating and superfluid phases~\cite{Greentree_NatPhys2006,Hartmann_NatPhys2006}, induced by the competition between photon blockade \cite{tian_quantum_1992, imamoglu_strongly_1997, rebic_large_1999} and inter-cavity photon tunneling has stimulated a great deal of discussion~\cite{hartmanLPR,fazio_review,hakan_nat_phys,jens,Angelakis_LL_PRL11}. More recently proposals appeared to realize artificial gauge fields~\cite{Koch_PRA11}, photonic lattices with non trivial band topology supporting chiral edge modes~\cite{Petrescu_PRA12,HafeziFanMigdallTaylor_arxiv2013} and quantum Hall fluids of light~\cite{Angelakis_FQH_PRL08,umucallar_artificial_2011,HafeziLukinTaylor_NJP2013}.

A major challenge on the way toward quantum simulation with photons arises due to the very nature by which photons interact with their environment and with the matter field.
In contrast to bosonic massive particles, photons can be annihilated by creating matter-like excitations or by exciting modes of their electromagnetic environment,
hence their number is not conserved. To describe this situation for a photon gas one says that photons have zero chemical potential~\cite{Landau5}. As a consequence, in order to engineer non-trivial quantum many body states other than the vacuum \cite{klaers_bose-einstein_2010} one would require an external non-equilibrium drive to balance the unavoidable losses. While the intrinsic non-equilibrium nature of these systems~\cite{AndrewNatPhys} represent a major source of excitement that fits nicely with the recent interest in the physics of strongly interacting quantum systems in different non-equilibrium regimes, the possibility to have non trivial equilibrium physics in a model of photons and atoms that could be potentially engineered in a lab represents an interesting perspective, worth to be pursued. 

In this paper, by going back to the basic elementary CQED Hamiltonian, the Rabi model, we investigate the equilibrium phase diagram of a lattice of CQED systems. In particular we will show that the inclusion of counter rotating terms in the light-matter interaction, often neglected in theoretical descriptions based on Jaynes-Cumming models, is crucial to stabilize finite-density quantum phases of correlated photons out of the vacuum, with no need for an artificially engineered chemical potential, the role of which can be effectively played by the light-matter interaction strength. As a result of an interaction term that explicitly breaks the conservation of total number of excitations, the quantum phases and the phase transition occurring between them turn to be remarkably different from those of massive bosonic particles. The competition between photon delocalization and Rabi non-linearity drives the system across a novel $Z_2$ parity symmetry-breaking quantum phase transition between two gapped phases, a Rabi insulator and a delocalized super-radiant phase where the TLSs polarize to generate a ferroelectrically ordered state and the photon coherence acquires a non-vanishing expectation value due to hopping. This novel quantum criticality shares some similarity with the Dicke phase transition of quantum optics, with the addition of non trivial dynamical and spatial quantum fluctuations, and turns out to be in the universality class of the Ising model.  We draw a complete picture of this novel quantum phase transition, that we have studied in a recent paper~\cite{Schiro_PRL12}, including the calculation of the response to a weak local driving which allows to probe the photon spectral function. We note that a similar model with Rabi non-linearity, although still in the presence of an artificial chemical potential, has been recently discussed in the literature~\cite{Zheng_PRA11}. The results of this study qualitatively agree with ours, even though their consequences on the universality class of the quantum phase transition have not been fully discussed.

The paper is organized as follows. In section ~\ref{sect:singlesite} we introduce the Rabi Model describing a single photonic mode coupled to a two level system. In section \ref{sect:latticerabi} we move to a tight-binding lattice model of coupled resonators with Rabi non-linearity. We discuss the nature of the quantum phase transition in this model, present the Gutzwiller mean field phase diagram and an effective low energy description in terms of an Ising spin model in a transverse field that allows to access the spectrum of excitations. Section \ref{sect:conclusions} is devoted to discussions and conclusions.

\section{Single Resonator Limit: The Rabi Model}\label{sect:singlesite}

We consider a single CQED unit with a two-level system (TLS) coupled to a radiation field. The basic Hamiltonian, introduced by Rabi~\cite{Rabi,haroche_exploring_2006}, reads
\be\label{eqn:HR}
\m{H}_{R}= \omega_0\,a^{\dagger}\,a+\omega_q\sigma^+\,\sigma^-+
g\,\left(a^{\dagger}+a\right)\,\left(\sigma^+ + \sigma^-\right)
\ee
In the following we introduce the detuning $\delta$ between the resonator and the TLS 
$$
\delta=\omega_q-\omega_0
$$
and mainly focus on the resonant case ($\delta=0$) unless explicitly stated. The light-matter interaction contains both rotating and counter-rotating terms which do not conserve the total number of excitations, $\m{N}=a^{\dagger}\,a+\sigma^+\sigma^-$. These are often discarded, within the so called rotating-wave approximation (RWA), to obtain the celebrated Jaynes Cumming (JC) Hamiltonian, that we write below for later convenience.
\be
 \m{H}_{JC}= \omega_0\,a^{\dagger}\,a+\omega_q\sigma^+\,\sigma^-+
g\,\left(a^{\dagger}\,\sigma^- + \sigma^+\,a\right)
\ee
 The presence of counter-rotating terms has some important consequences, that we are now going to discuss. The $U(1)$ symmetry associated with conservation of number of excitations breaks down to a discrete $Z_2$ symmetry generated by the parity operator $\m{P}$ defined as
\be\label{eqn:parity}
\m{P}=e^{i\pi\,\m{N}}\,\qquad\,\m{N}=a^{\dagger}a+\spin{+}\spin{-}=a^{\dagger}a+\frac{1}{2}\,(\sigma_z+1)
\ee
It is easy to see that the action of the parity on the photonic degrees of freedom is
\be
\m{P}^{\dagger}\,a\,\m{P}=-a \,\qquad
\m{P}^{\dagger}\,a^{\dagger}\,\m{P}=-a^{\dagger}
\ee
while for the two level system we have
\be
\m{P}^{\dagger}\,\sigma_x\,\m{P}=-\sigma_x \,\qquad
\m{P}^{\dagger}\,\sigma_z\,\m{P}=\sigma_z \,.
\ee
We therefore conclude that the action of the parity operator leaves the Rabi Hamiltonian invariant 
\be
\m{P}^{\dagger}\,\m{H}_{R}\,\m{P}= \m{H}_{R}\,\qquad\,
\longrightarrow\,\qquad[\m{P},\m{H}_{R}]=0
\ee
While the discrete $Z_2$ symmetry prevents a full closed-form solution, the Rabi model~(\ref{eqn:HR}) has recently been shown to be nevertheless integrable~\cite{Braak_PRL11} and its spectrum written in implicit form. In order to get more physical intuition into the nature of this solution, we proceed here by exact numerical diagonalization of the  Rabi model by introducing a cut-off on the maximum number of photons $N_{max}$ to truncate its Hilbert space.
\begin{figure}[t]
\begin{center}
\epsfig{figure=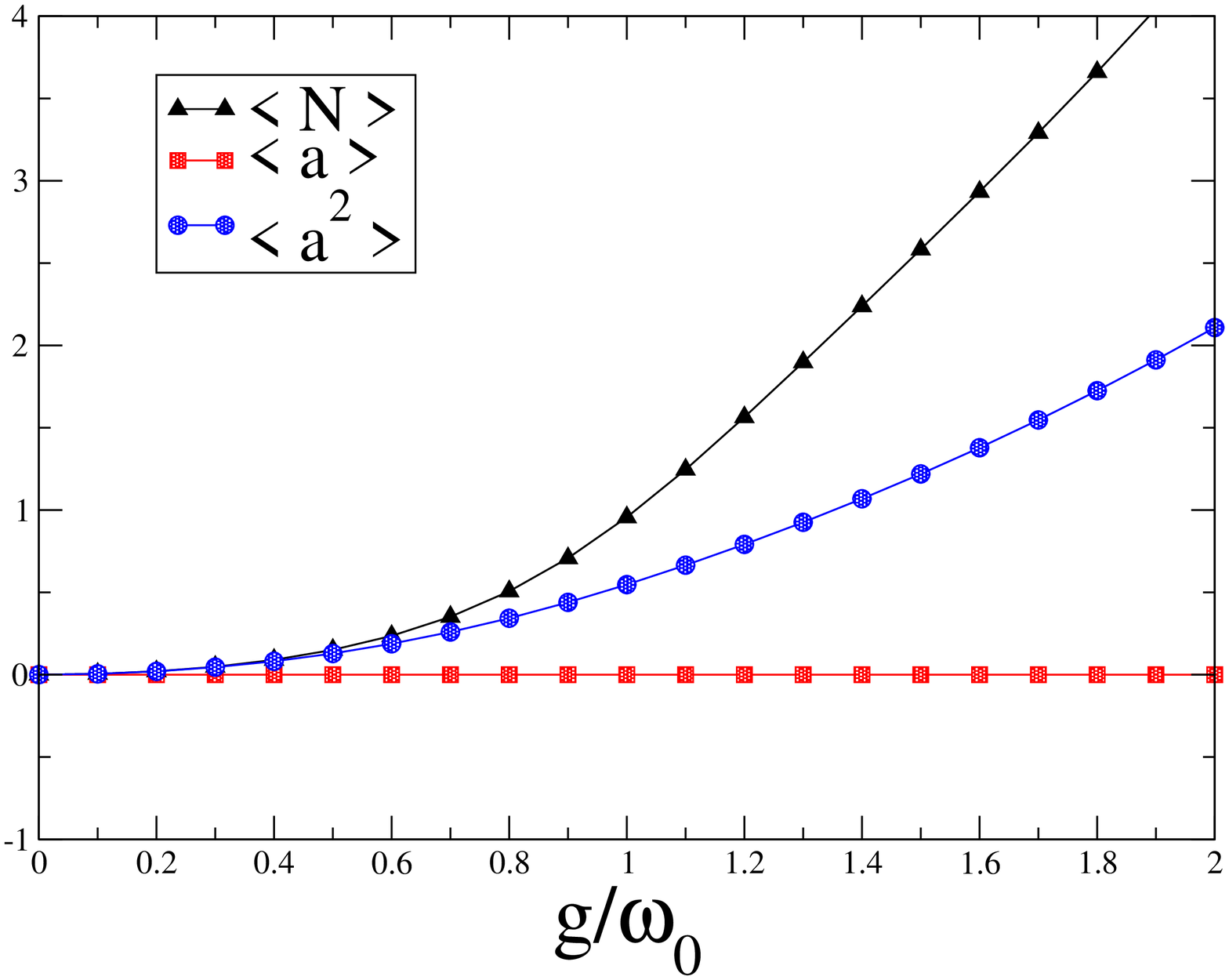,scale=0.3}
\caption{Ground state properties of the single site Rabi Model.  We plot the average number of excitations in the ground state $\langle\,\m{N}\rangle$, the average value of the photon field in the resonator $\langle\,a\rangle$ and the average squeezing $\langle\,a^2\rangle$ as a function of $g/\omega_0$, at zero detuning $\delta=0$.} 
\label{fig:singlesite_rabigs}
\end{center}
\end{figure}

Let us now come back to the role of the $Z_2$ symmetry. As the number of excitations is not conserved the exact eigenstates of the Rabi model will be labelled only by the parity quantum number (in addition to energy). This immediately implies, by symmetry, that the ground state of the single site Rabi model will have, for any $g/\omega_0$,
\be
\langle a\rangle = 0\,\qquad\,
\langle \sigma_x\rangle = 0
\ee
since both operators are odd under $Z_2$. In addition, as the ground state will be a linear combination of states with different number of excitations, the average number of photons in the Rabi groundstate will be different from zero and, more specifically, a smoothly increasing function of $g/\omega_0$, even in absence of an external driving. This is indeed confirmed by the exact numerical solution of the model. In figure \ref{fig:singlesite_rabigs} we plot the average number of excitations $\langle\,\m{N}\rangle$ and the average value of the photon operator $\langle\,a\rangle$ (the TLS $\langle\,\sigma_x\rangle$ behaves similarly) as a function of $g/\omega_0$. We notice the former increases smoothly with $g/\omega_0$. Similarly, the order parameters of the $Z_2$ symmetry are zero, as the ground state has always a well defined parity. 
\begin{figure}[t]
\begin{center}
\epsfig{figure=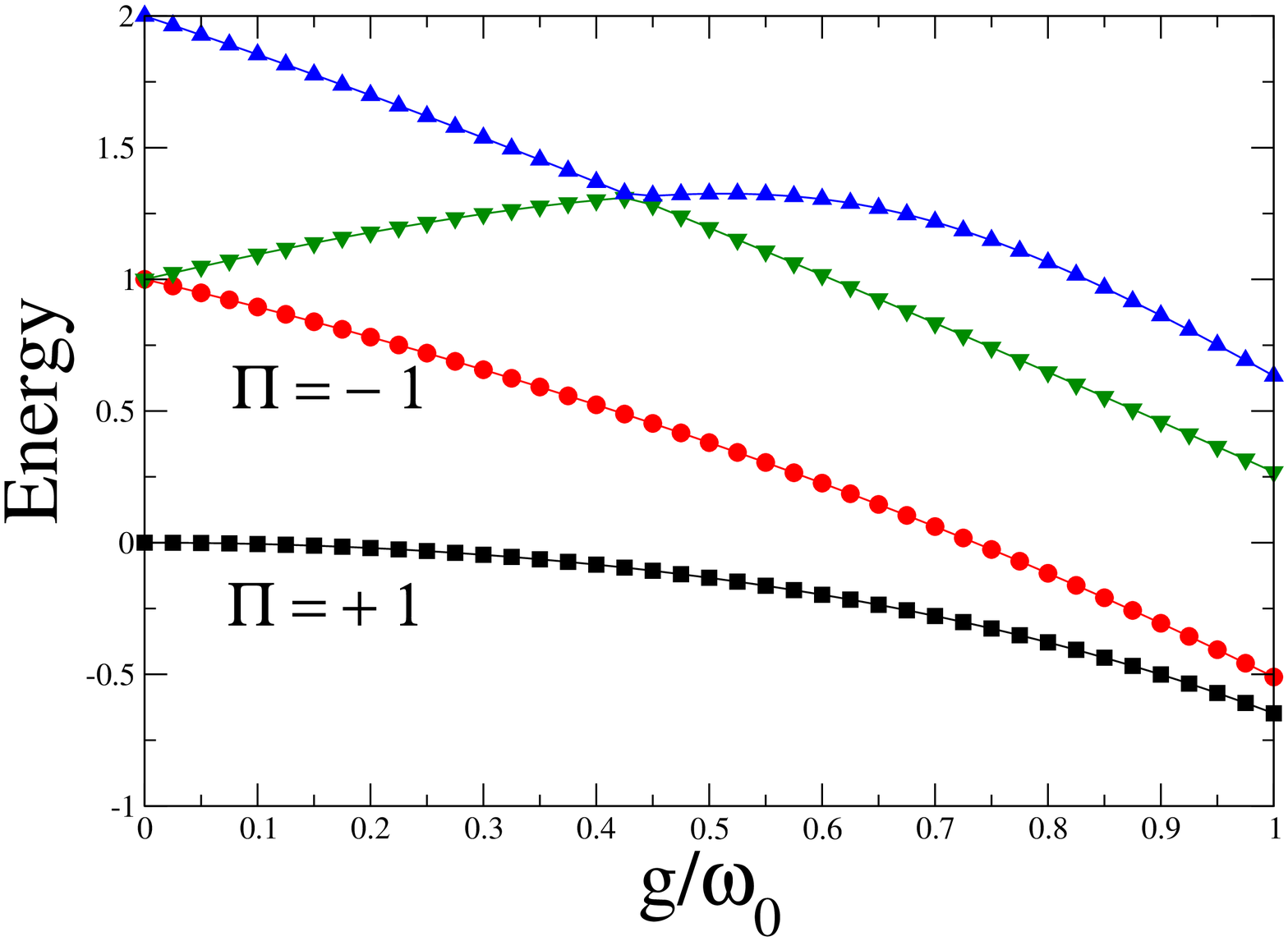,scale=0.3}
\caption{Low-lying spectrum of the single site Rabi Model. We plot as a function of the light matter interaction $g/\omega_0$ and at zero detuning $\delta=0$ the ground state and the first few excited states, labelled by the parity $\Pi=\pm$. We notice that the ground-state and the first excited state are almost degenerate for large $g/\omega_0$ and that the former has always a fixed parity.} \label{fig:rabi_spectrum}
\label{fig:Npol_rabi}
\end{center}
\end{figure}
This comes with no surprises, as the single site Rabi model involves just a single atomic and a photonic mode that cannot result in symmetry breaking (the model is zero dimensional). It is worth in this respect to stress the difference with the more often discussed Dicke model~\cite{baumann_dicke_2010}, where a large number $N$ of TLS are coupled to a single photonic mode and features in the thermodynamic limit  $N\rightarrow\infty$ a genuine spontaneous symmetry breaking of the parity. 

In addition to being populated by a finite number of excitations and symmetric, the ground state of the Rabi Hamiltonian is also squeezed, i.e. $\langle\,a^2\rangle\neq0$. In particular, due to the discrete nature of the parity symmetry, one can have for the Rabi ground state that 
$\langle\,\left(a\right)^{2m}\rangle\neq0 $
since by creating an even number of photons the overall parity is not affected.  This property will play an important role when discussing the critical behavior of the lattice Rabi model.
%At this point a comment on a related issue may be useful in connection with the physics of the Dicke Model where a single mode of the photon field interact with $N$ two level systems. In that case a spontaneous breaking of the parity symmetry arises, in the thermodynamic limit $N\rightarrow\infty$. Here we are dealing with a zero dimensional model where no spontaneous symmetry breaking can arise.
\begin{figure}[h]
\begin{center}
\epsfig{figure=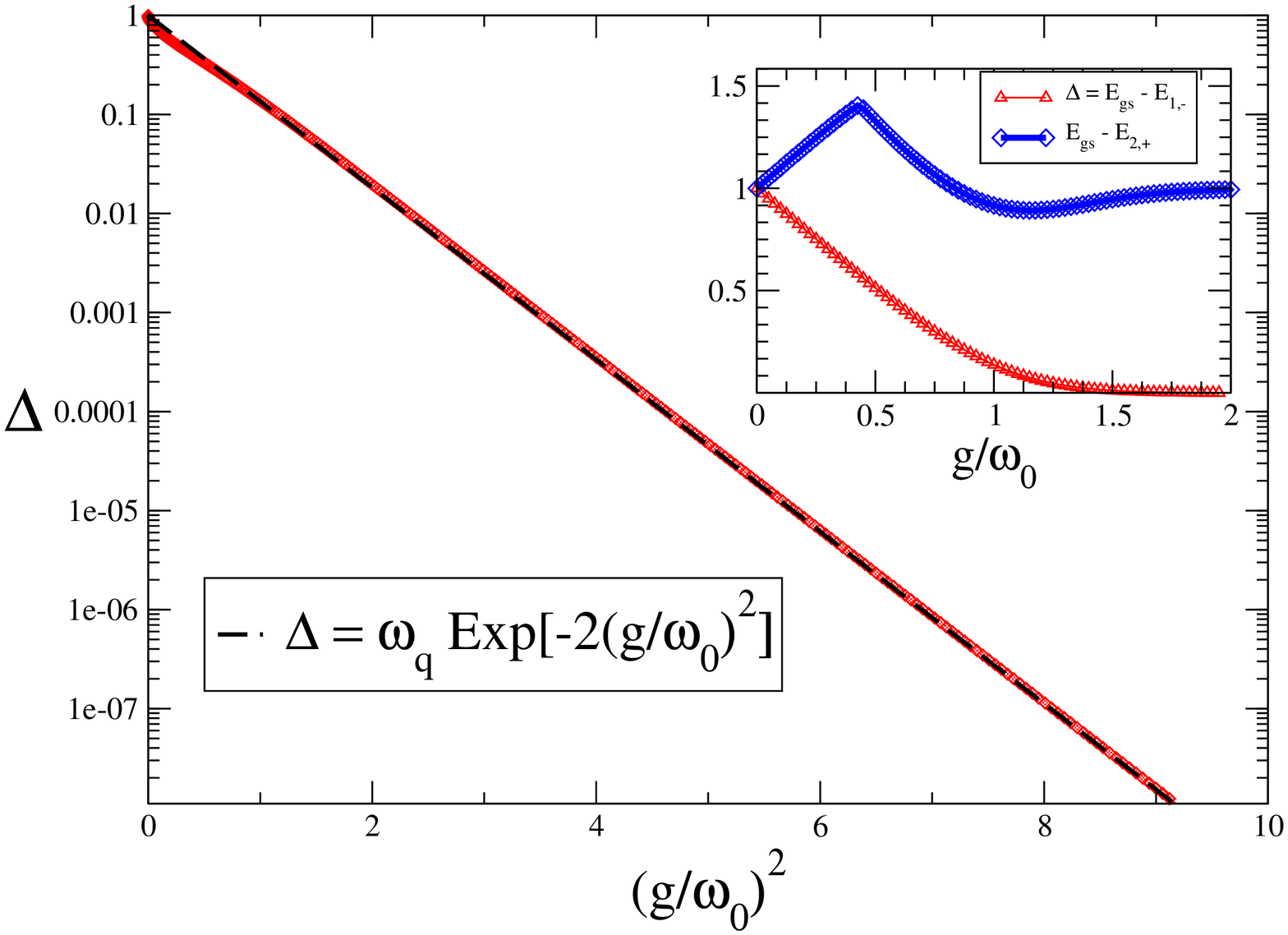,scale=0.3}
\caption{Splitting between the ground-state and the first excited state in the single site Rabi model. The red line is the exact numerical solution, the dashed black line is the asymptotic expression $\Delta=\omega_q\,e^{-2(g/\omega_0)^2}$. In the inset we plot the gap to the second excited state, which remains of order one even at large coupling $g\gg\omega_0$. } 
\label{fig:splitting}
\end{center}
\end{figure}
We now discuss the structure of the low lying excited states of the Rabi model.  In figure ~\ref{fig:rabi_spectrum}, we plot the four lowest energy levels of the Rabi model at zero detuning as a function of $g/\omega_0$. We notice, quite interestingly, that (i) the ground state of the Rabi model remains an even parity state for any $g/\omo$, i.e. no level crossing between the ground state and the first excited state occurs as a function of $g/\omo$, and (ii) the ground state and the first excited state are quasi-degenerate in the ultra-strong coupling regime $g\gtrsim\omo$ with an exponentially small energy splitting $\Delta$, while the gap to the next energy level stays of order one (see \Fig{fig:splitting}). Since this latter observation will play a crucial role later on in constructing an effective low energy model for the Rabi lattice model,  in the next subsection we give a simple analytical explanation for the exponential splitting that agrees well with the numerical result.

\subsection{Splitting in the ultrastrong coupling regime}

Here we want to discuss the single site Rabi model in the limit $g/\omega_0\gg1$. From the exact solution we notice that the ground state and the first excited state are almost degenerate in this ultra strong coupling regime (see figure~\ref{fig:rabi_spectrum}) and with opposite parity. Here we want to get additional insight into this quasi-degeneracy and in particular to compute the splitting $\Delta$ as a function of $g/\omega_0$.
Let us start from the Rabi Hamiltonian that we rewrite for simplicity here
\be
\m{H}= 
\omega_0\,a^{\dagger}\,a+\frac{\omega_q}{2}\,\sigma_z+
g\,\left(a^{\dagger}+a\right)\,\sigma_x
\ee
For $\omega_q=0$ the model can be trivially solved. Indeed in this case $\sigma_x$ is conserved, namely we can define two Hamiltonians
\be
\m{H}_{\pm}= \omega_0\,a^{\dagger}\,a\pm g\,\left(a^{\dagger}+a\right)\,
\ee
representing two displaced harmonic oscillators. Let us define as $\varphi_{\pm}(x)$ the ground states of these two Hamiltonians, corresponding respectively to $\pm g$. It is clear that the ground state of the original $\m{H}$ for $\omega_q=0$ is doubly degenerate as the two states
\be 
\vert\Psi_{\pm}\rangle = \varphi_{\pm}(x)\vert\sigma_x\rangle\,\qquad\,
\vert\sigma_x\rangle=\vert\rightarrow\rangle,\vert\leftarrow\rangle
\ee 
are exactly degenerate. Indeed the groundstate energy of the displaced harmonic oscillator reads
\be
E_{\pm}=\frac{\omega_0}{2}-\frac{g^2}{\omega_0}
\ee
independent of the sign of $g$. Using the two above degenerate states we can construct two new states with well defined parity:
\bea 
\vert\pm\rangle_0 =\frac{1}{\sqrt{2}}\,
\left(\varphi_+(x)\vert\rightarrow\rangle\pm
\varphi_-(x)\vert\leftarrow\rangle
\right)
\eea
It is easy to see that under a parity transformation
\be
\m{P}\vert\,+\rangle_0=\vert\,+\rangle_0\,\qquad\,
\m{P}\vert\,-\rangle_0=-\vert\,-\rangle_0\,\qquad
\ee
Any finite $\omega_q$ gives quantum mechanical fluctuations to $\sigma_x$ and results into a lifting of this degeneracy. The perturbation 
\be
V_{\sigma^z}=\frac{\omega_q}{2}\,\sigma_z 
\ee
results in a splitting
\bea
\Delta =  \frac{\omega_q}{2}\,\left[\langle\,+\vert\,\sigma_z\vert\,+\rangle_0-
\langle\,-\vert\,\sigma_z\vert\,-\rangle_0
\right]=\nonumber\\
=\omega_q\,\int\,dx\,\varphi_+(x)\varphi_-(x)=\omega_q\, e^{-2g^2/\omega_0^2}
\eea
The agreement between this result and the exact numerics in the strong coupling regime $g/\omega_0\gg1$ is remarkable as shown in figure~\ref{fig:splitting}. 

\section{The Rabi Hubbard Model}\label{sect:latticerabi}

We now consider a model of localized photonic modes, $a_{\bR},a^{\dagger}_{\bR}$, hopping at a rate $J$ on a lattice with coordination $z$, and locally coupled by a Rabi interaction~(\ref{eqn:HR}) to a set of a TLSs, described by Pauli operators $\sigma^{\pm}_{\bR}$. The Hamiltonian for this interacting photonic lattice model that we refer to as the Rabi-Hubbard (RH) model reads
\be\label{eqn:HlatticeRabi}
\m{H }= -\sum_{\langle\,\bR\bRp\rangle}\,\,J_{\bR\bRp}\,
x_{\bR}\,x_{\bRp} + \sum_{\bR}\,H_{R}[a_{\bR},a^{\dagger}_{\bR},\sigma^{x}_{\bR}]\,
\ee
where $x_{\bR}=a_{\bR}+a^{\dagger}_{\bR}$ while the single site local Hamiltonian reads as
\bea
H_{R}[a,a^{\dagger},\sigma^{\pm}]=
 \omega_0\,a^{\dagger}\,a+\omega_q\sigma^+\,\sigma^-\nonumber+
 g\left(a+a^{\dagger}\right)\,\sigma^x
\eea
We note that in the above Hamiltonian we have included the counter-rotating terms in the hopping for the photon field at neighboring sites.
% that arises in principle from the capacitive or inductive coupling between resonators.
%Few comments are in order at this point. First, we notice that in light of possible future experiments on circuit QED architectures, we do not include any chemical potential to tune the density of excitations in the ground state. Rather we would like here to exploit the spontaneous polarization of the Rabi vacuum that emerges when the light-matter interaction strength $g$ is increased.
%In addition, we notice that in principle the capacitive or inductive coupling between resonators may also induce counter-rotating terms for the photon field at neighboring sites.  In order not to deal explicitly with those terms we limit ourself here to the regime of weak coupling between resonators, where a rotating-wave approximation on the hopping hamiltonian might well be justified. We notice however that 
%having included the counter-rotating terms explicitly in the light-matter interaction we do not expect those anomalous hopping terms to qualitatively alter the physics of the Rabi Hubbard lattice model in terms of its universal properties and critical behaviour.

Two trivial limits of the RH Hamiltonian~(\ref{eqn:HlatticeRabi}) can be immediately discussed. In the limit vanishing hopping, $J_{\bR\bRp}=0$ all lattice sites decouple and the Hamiltonian reduces to a collection of single resonators with Rabi non-linearity, whose physics we have discussed in the previous sections. In the opposite limit of vanishing non-linearity $g=0$ the photonic modes delocalize throughout the lattice and form a band. In the generic case $J\neq0,g\neq0$ the model cannot be solved exactly and a non trivial behaviour is expected to emerge out of the competition between photon delocalization and Rabi non-linearity~\cite{Schiro_PRL12}.

The roots of this can be traced back to the single resonator limit. As discussed above, the counter-rotating terms leave the system with a discrete $Z_2$ symmetry associated to parity. Photon hopping in (\ref{eqn:HlatticeRabi}) can trigger a spontaneous breaking of this parity symmetry above some critical coupling $J_c(g)$, toward a phase where both $\langle\,a_\bR\rangle\neq0$ and $\langle\,\sigma^x_\bR\,\rangle\neq0$. As we are now going to show the Rabi Hubbard model belongs to the Ising universality class and thus is fundamentally different from the Jaynes-Cumming one (see \cite{jens} and references therein). Instead, it can be seen as a delocalized super-radiant critical point reminiscent of the Dicke transition of quantum optics, an open version of which was recently realized, effectively, by coupling motional degrees of freedom of an atomic BEC to a single mode of an optical cavity~\cite{baumann_dicke_2010}. In contrast to the single-mode Dicke model however, which is an exact mean-field phase-transition, the Rabi-Hubbard model displays non-trivial spatial and dynamical fluctuations. In the following we will discuss the qualitative physics of the Rabi Hubbard model and its phase diagram using different analytical approaches.

\subsection{Mean Field Theory}\label{sect:meanfield}

Let's start deriving a mean field theory for the Rabi Hubbard model. Following the standard approach for bosons~\cite{Fisher_Fisher_PRB89} we decouple the hopping term in~(\ref{eqn:HlatticeRabi}) on a bond $\langle\,\bR\bRp\rangle$ as
\be
H_{hop}= x_{\bR}\,x_{\bRp}\simeq\,x_{\bR}\,\langle\,x_{\bRp}\rangle+\langle\,x_{\bR}\rangle\,x_{\bRp}-
\langle\,x_{\bRp}\rangle\,\langle\,x_{\bRp}\rangle
\ee
so to reduce the original lattice Hamiltonian~(\ref{eqn:HlatticeRabi}) to an effective single site problem 
\be 
\m{H}\simeq  \sum_{\bR}\,H^{eff}_{\bR}[a_{\bR},a^{\dagger}_{\bR},\sigma^x_{\bR}]
\ee
with an effective Hamiltonian
\bea
H^{eff}_{\bR}=\omega_r\,a^{\dagger}_{\bR}\,a_{\bR}+\omega_q\,\sigma^+_{\bR}\sigma^-_{\bR}  +g\,
\left(a_{\bR}+a^{\dagger}_{\bR}\right)\,\sigma^x_{\bR}+\nonumber\\
+\psi_{\bR}\left(a_{\bR}+a^{\dagger}_{\bR}\right)
\eea
describing a single site Rabi model in a coherent driving field $\psi_{\bR}$ that has to be self-consistently determined according to
\be\label{eqn:meanfield_sc}
 \psi_{\bR} = -\sum_{\bRp\in v(\bR)}\,J_{\bR\bRp}\langle\,x_{\bRp}\rangle_{\psi_{\bRp}}
\ee
Alternatively the effective  field $\psi_{\bR}$ can determined by minimizing the total energy (at zero temperature)
\be\label{eqn:Egs}
E[\psi_{\bR}]=\sum_{\bR}\,\langle\,H^{eff}_{\bR}\rangle_{\psi_{\bR}}+
\sum_{\bR\bRp}\,J_{\bR\bRp}\,\psi_{\bR}\,\psi_{\bRp} 
\ee
We notice the above mean field approach is \emph{exact} for infinite coordination lattices, i.e. in the limit $z\rightarrow\infty$, and can be seen as the first step of a systematic expansion in $1/z$. Quantum fluctuations beyond this Gutzwiller solution can be accounted for through the Bosonic Dynamical Mean Field Theory~\cite{Pollet_prl11} or the Quantum Cavity Method~\cite{Zamponi_QuantumCavity_prb08}.

\begin{figure}[t]
\begin{center}
\epsfig{figure=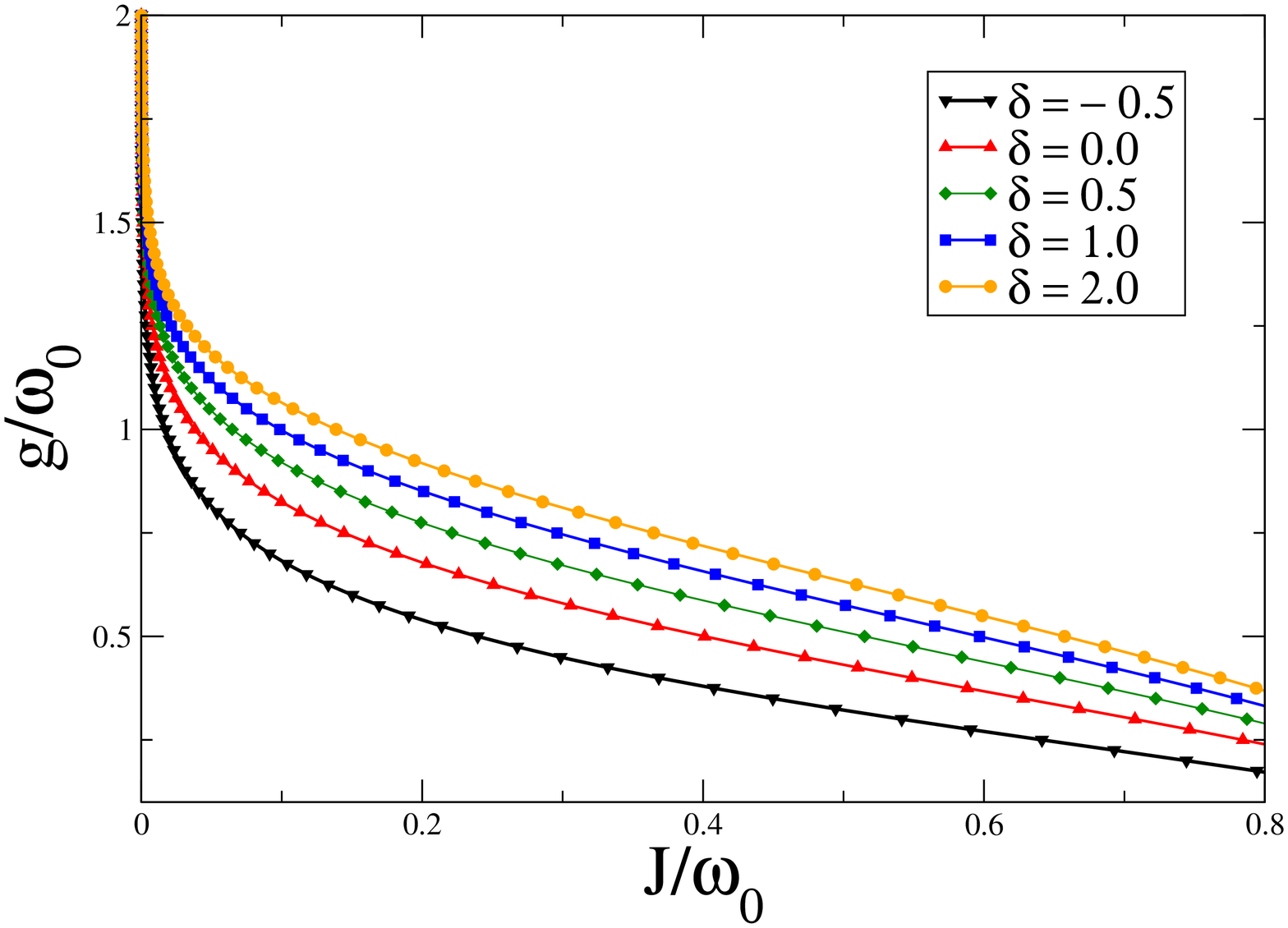,scale=0.3}
\caption{Mean Field Phase Diagram for the Rabi Hubbard lattice model. We plot the critical value of the light matter interaction $g$ as a function of the hopping strength $J$, at different values of detuning $\delta=(\omega_q-\omega_0)/\omega_0$. Above a certain critical hopping $J_c(g,\delta)$ the $Z_2$ parity symmetry is spontaneously broken.} 
\label{fig:meanfield_phasediag}
\end{center}
\end{figure}
%As we will discuss more extensively in section~\ref{sect:generalized_rabi}, we notice that only the first term would be non vanishing in absence of counter-rotating terms such as in the the Jaynes Cumming lattice model~\cite{Koch_LeHur_2009}, while in the Rabi Hubbard case they all contribute. 
The mean field phase boundary can be obtained, in the spirit of a Landau theory, by expanding the ground state energy~(\ref{eqn:Egs}) around the symmetric solution $\psi_{\bR}=0$. The phase transition toward a broken symmetry phase arises when the coefficient of the second order term becomes zero. This gives us the condition for the critical hopping $J_c$
\be
 \frac{1}{zJ_c}=\frac{1}{2}\,\,\int_0^\beta\,d\tau\,\langle\,T_{\tau}\,x(\tau)\,x(0)\rangle_{loc}=
\frac{1}{2}\,\,C_{xx}(i\Omega=0)
\ee
%From the definition of the effective potential $\Gamma[\psi]$ we get
%\bea
%\frac{\partial^2\,\Gamma}{\partial\psi^2}\vert_{\psi=0}=
%\int_0^\beta\,d\tau_1\,\int_0^\beta\,d\tau_2\,
%\left[\langle\,T_{\tau}\,X(\tau_1)\,X(\tau_2)\rangle_{loc}-\langle\,T_{\tau}\,X(\tau_1)\rangle_{loc}\,
%\langle\,T_{\tau}\,X(\tau_2)\rangle_{loc}\right]=\\
%=\int_0^\beta\,d\tau_1\,\int_0^\beta\,d\tau_2\,
%\langle\,T_{\tau}\,X(\tau_1)\,X(\tau_2)\rangle_{loc}=
%\int_0^\beta\,d\tau\,\langle\,T_{\tau}\,X(\tau)\,X(0)\rangle_{loc}=
%C_{XX}(i\Omega\rightarrow0)
%\eea
where we have defined the local bosonic correlation function in Matsubara frequencies as
\be
C_{xx}(i\Omega)= \int_0^\beta\,d\tau\,e^{i\Omega\tau}\langle\,T_{\tau}\,x(\tau)\,x(0)\rangle_{loc}
\ee
%\begin{figure}[t]
%\begin{center}
%\epsfig{figure=../Plot/EPS/energy_jc.eps,scale=0.3}
%\epsfig{figure=../Plot/EPS/psi_star_jc.eps,scale=0.3}
%\caption{Lattice Jaynes-Cumming Model, Gutzwiller Mean Field Theory. We work at zero detuning $\delta=0$ and with $g/\Omega=1.1$ (namely in the $n=1$ Mott lobe). Left Panel: Energy as a function of $\psi$ for different values of $\kappa_{eff}=ZJ=0.05,0.1,0.25,0.5$ from top to bottom. At a critical value of $\kappa_{eff}$ the system develops a non vanishing $\psi=\langle\,a\rangle$ which signals a phase transition into a broken symmetry phase. Right panel: We plot the behaviour of the order parameter $\psi=\langle\,a\rangle$ as a function of $\kappa_{eff}$.} 
%\label{fig:energy_mf_jc}
%\end{center}
%\end{figure}
Using a representation in terms of exact eigenstates of the single site Rabi Hamiltonian 
\be 
\m{H}_{R}\vert\,n\rangle=E_n\vert\,n\rangle
\ee
we can write the expression for the mean field  phase boundary at zero temperature as
\be
\frac{1}{zJ_c}=\sum_{n}\,
\vert\langle\,n\vert\,x\vert\,GS\rangle\vert^2\,
\frac{1}{E_n-E_{gs}}
\ee
where $\vert\,GS\rangle$ is the ground state of the single site Rabi model with energy $E_{gs}$.
In  \Fig{fig:meanfield_phasediag} we plot the mean field phase boundary in the $J,g$ plane for different values of the detuning $\delta$. We see that quite generically a disordered phase is stable until a critical hopping strength $J_c(g,\delta)$ is reached above which a broken symmetry phase emerges. This is characterized by the photon field becoming locally coherent, $\langle\,a_{\bR}\rangle\neq0$, and the TLS ordering $\langle\,\sigma_{\bR}^x\rangle\neq0$. Upon increasing the relative detuning between the TLS and the photon, $\delta=(\omega_q-\omega_0)/\omega_0$, from negative toward positive values the former will tend to fluctuate more strongly, as a consequence of a larger local transverse field $\omega_q\sigma^z$. This will favour local disordering, hence one may expect the size of the broken symmetry ordered phase to shrink, as indeed confirmed in Fig.~\ref{fig:meanfield_phasediag}.

Due to the discrete nature of the symmetry involved, a $Z_2$ associated to the parity, the qualitative features of the two phases and the transition between them are expected to be different, in terms of universal behaviour and physical properties, from the more often discussed Jaynes-Cumming lattice model~\cite{hohenadler_dynamical_2011} with its superfluid to Mott transition of polaritons. 
In particular, the $Z_2$ symmetry of the order parameter points toward an Ising universality class of the quantum phase transition separating two gapped phases, with a gap vanishing as a power-law at the critical point. In the next section we will confirm this scenario by introducing an effective low-energy theory for the Rabi-Hubbard model. This will allow us to compute the spectrum of low-lying excitations across the phase diagram and to evaluate the photon Green's function by including gaussian fluctuations around the mean field theory.

\subsection{Effective Low Energy Description: Ising Model in a Transverse Field}

Here we discuss an effective low energy description for the lattice Rabi model, which is valid asymptotically in the regime $g\gg\omega_0$. In this case we know from the exact solution of the single site problem (see section~\ref{sect:singlesite}) that the groundstate and the first excited state are almost degenerate and with opposite parity. This suggests to represent this low energy doublet $\vert\pm\rangle$ as eigenstates of an effective local degree of freedom, a pseudo-spin $1/2$ living on each site $\bR$ of the lattice
\be 
\Sigma^z_{\bR}\,\vert\pm\rangle=\pm\,\vert\pm\rangle
\ee
Since the gap to the next energy level in the spectrum of the single site Rabi model stays of order one at large $g/\omo$,  this suggests building an effective Hamiltonian by projecting the local Hilbert space onto this low energy doublet  $\vert\pm\rangle$. The local Hamiltonian in this truncated basis reads, by definition
\be
\m{H}_{R}\simeq E_+\vert\,+\rangle\langle\,+\vert+ E_-\vert\,-\rangle\langle\,-\vert=
\frac{\bar{E}}{2}\,\,\mathbb{I}+
\frac{\Delta}{2}\,\Sigma^z
\ee
where $E_{\pm}$ are the two lowest eigenvalues of the Rabi spectrum, $\Delta= E_+-E_->0$ the level splitting and $\bar{E}$ the average energy between the two levels. The latter contributes an irrelevant identity matrix that will be dropped in the following. Written in the pseudo-spin basis $\vert\pm\rangle$, the photon creation/annihilation $a_{\bR}, a^{\dagger}_{\bR}$ take a purely off-diagonal form because they do not couple states with the same parity.
%In order to build an effective Hamiltonian we need to write the photon creation/annihilation operators $a_{\bR},a^{\dagger}_{\bR}$ in this new basis $\vert\pm\rangle$. We notice in this respect that by symmetry the photon operator $a_{\bR}$ does not couple states with same parity, hence in the $\vert\pm\rangle$ basis it will take a purely off diagonal form (with $\beta\neq\gamma$)
$$
\langle\,n\vert\,a\,\vert\,m\rangle = \left(
\begin{array}{ll}
0 & \beta\\
\gamma & 0
\end{array}
\right)=\beta\,\Sigma^++\gamma\,\Sigma^-
$$
Similarly we get for the hermitian conjugate $a^{\dagger}$
$$
\langle\,n\vert\,a^{\dagger}\,\vert\,m\rangle = \left(
\begin{array}{ll}
0 & \gamma\\
\beta & 0
\end{array}
\right)=\gamma\,\Sigma^++\beta\,\Sigma^-
$$
Using these results we can now write the photon hopping as an effective spin-spin interaction between neighboring sites. A straightforward calculation gives the following
effective spin model Hamiltonian 
\be\label{eqn:Heff}
\m{H}_{eff} = -\sum_{\langle\bR\bR'\rangle}\,
J_x\,\Sigma^x_{\bR}\,\Sigma^x_{\bRp}
%\left(J_x\,\Sigma^x_{\bR}\,\Sigma^x_{\bRp}+
%J_y\,\Sigma^y_{\bR}\,\Sigma^y_{\bRp}\right)
+\frac{\Delta}{2}\sum_{\bR}\,\Sigma_\bR^z
\ee
with $J_x =\frac{J}{2}\left(\gamma+\beta\right)^2$, 
%\be
%J_x = \frac{J}{2}\left(\gamma+\beta\right)^2\,\qquad\,
%J_y = \frac{J}{2}\left(\gamma-\beta\right)^2\,\qquad\, 
%\ee
where $J$ is the hopping between the photonic resonators in the original Rabi Hubbard model~(\ref{eqn:HlatticeRabi}), that we take here between neighboring sites only. The resulting Hamiltonian describes an Ising spin model in a transverse magnetic field  which is known to display a quantum phase transition toward a $Z_2$ broken symmetry phase which is in the Ising universality class~\cite{sachdev_quantum_2011}. We note that the presence of a finite asymmetry between the rotating and counter-rotating terms in the photon hopping term would lead to an anisotropic spin-exchange $J_x\neq J_y$ which remains within the same Ising universality class. 
% We notice that in presence of a finite asymmetry between the rate of rotating and counter-rotating hopping terms in the original photon hamiltonian, that we have taken here to be equal as it is natural in actual experimental implementations, would have result into a finite spin exchange along the $y$ direction, $J_x\neq J_y$, with no qualitative change of the physical picture we are going to draw~\cite{Schiro_PRL12}. 
The parameters entering the effective model, $J_{x},\Delta$, depend strongly on the single-site physics and can be obtained numerically from the exact diagonalization of the Rabi Hamiltonian $H_{R}$.
However their behaviour at strong light matter coupling,  $g\gg\omega_0$, can be obtained analytically. Indeed from the analysis of section~\ref{sect:singlesite} the level splitting was found to be exponentially suppressed at strong coupling, $\Delta=\omega_0\,\exp\left[-2\left(g/\omega_0\right)^2\right]$. Similarly,  one can write the photon field operator $a$ in the basis of the low-energy doublet $\vert\pm\rangle$. To leading order we get
for the off-diagonal terms $
\langle\,-\vert\,a\,\vert+\rangle=\langle\,+\vert\,a\,\vert-\rangle\simeq g/\omega_0$
from which we conclude that for $g\gg\omega_0$ we have $J_x\simeq 2J\left(g/\omega_0\right)^2$ which we find to nicely match the numerical results obtained by numerical diagonalization~\cite{Schiro_PRL12}.

\subsubsection{Low Energy Spectrum: Mean Field and Fluctuations}

In this section we analyze the low-energy excitations of the effective pseudospin-model Eq.~(\ref{eqn:Heff}) through the spin-wave theory.
%To get further insights into the nature of the excitation spectrum of the Rabi Hubbard model we now study the effective spin Hamiltonian using a mean field plus fluctuations approach. In the spirit of a spin-wave theory we first look for the classical ordered ground state and then derive an harmonic theory for the fluctuations around it. 
The classical ground state can be obtained by rotating each spin locally around $\Sigma^y$ with a unitary operator
\be\label{eqn:rotation}
 \Omega = \exp\left(i\theta\Sigma^y/2\right)
\ee
The angle $\theta$ is then fixed by minimizing the energy of a classical ferromagnetic state along $x$ in the rotated basis, which gives the condition
\be\label{eqn:minimum}
\frac{\partial\,E_{cl}}{\partial\theta}=cos\theta\left(zJ_x\,sin\theta\,+\frac{\Delta}{2}\right)=0 
\ee
This identifies two regimes, namely for $2zJ_x>\Delta$ we have
\be\label{eqn:beta_mf}
 sin\theta = - \frac{\Delta}{2z\,J_x}\,
 \ee
while for $2zJ_x<\Delta$ we have  $\sin\theta=-1$, $\cos\theta=0$. When expressed in terms of the hopping $J$ between photons this identifies the mean field phase boundary 
 \be
z J_c = \frac{\Delta}{\left(\beta+\gamma\right)^2} 
 \ee
% In the following we will choose this as unit of energy and introduce the reduced hopping $\m{J}=J/J_c$.
If we compute the average magnetization along $x$ we find for $J>J_c$
\bea
 \langle\Sigma^x_{\bR}\rangle=cos\theta=\sqrt{1-\left(J_c/J\right)^2}
%\langle\,\Sigma^z_\bR\rangle=sin\theta =-J_c/J
\eea
while $ \langle\Sigma^x_{\bR}\rangle=0$ for $J<J_c$ from which we can conclude that at the classical level there is an Ising like phase transition at $J_c$ between a quantum disordered paramagnet for $J<J_c$ and a ferromagnetically ordered state for $J>J_c$.
  
Once we have found the classically ordered groundstate playing the role of the vacuum $\vert0\rangle$ we can now include the quantum fluctuations around this state at the leading (harmonic) order within a spin wave approach. To do that we introduce the following representation of $\Sigma_{\bR}^{\alpha}$ operators in terms of a set of harmonic oscillators
\bea\label{eqn:bosonx}
\Sigma^x_{\bR}=1-\Pi_{\bR}\\
\label{eqn:bosony}
\Sigma^y_{\bR}=-\sqrt{2}\,p_\bR\\
\label{eqn:bosonz}
\Sigma^z_{\bR}=\sqrt{2}\,x_{\bR}\,
\eea
where $\Pi_{\bR}=x^2_{\bR}+p_{\bR}^2-1$ measure the strength of the quantum fluctuations, namely
\be 
\langle\,\Pi_{\bR}\rangle=0
\ee
when evaluated on the vacuum (i.e. the classical ground state).
% It is easy to see that the above bosonic representation satisfies the spin algebra
%\be
%[\Sigma^a_{\bR},\Sigma^b_{\bR}]=i\eps_{abc}\,\Sigma^c_{\bR} 
%\ee
We now substitute the expressions~(\ref{eqn:bosonx}-\ref{eqn:bosonz}) in the effective Hamiltonian  $\bar{\m{H}}_{eff}$ and keep only terms up to linear order in $\Pi$. Then, by using the condition~(\ref{eqn:minimum}) to cancel terms which are linear in the oscillators, we end up with a collection of harmonic oscillators describing the low lying fluctuating modes around the classical ground state. The resulting Hamiltonian is then readily diagonalized in terms of bosonic modes $b_{\bq},b^{\dagger}_{\bq}$
\be
\bar{H}_{qf}
%=\sum_{\bq}\,A_{\bq}\,x_{\bq}\,x_{-\bq}+B_{\bq}\,p_{\bq}\,p_{-\bq}
=
\sum_{\bq}\,\omega_{\bq}\,b^{\dagger}_{\bq}\,b_{\bq}
\ee
where $\omega_{\bq}$ describes the spectrum of low-lying excitations whose structure we now discuss as the hopping $J$ is tuned across the critical point.
In the disordered phase $J<J_c$ we find
\be
\omega_{\bq}=
\Delta\,\sqrt{\left(1-\m{J}\gamma_{\bq}\right)}
%\left(1-\frac{J}{J_c}\xi\gamma_{\bq}\right)}
\ee
where we have introduced the reduced hopping $\m{J}=J/J_c$,  while for $J>J_c$ we get
\be
\omega_{\bq}=
z\,J\left(\beta+\gamma\right)^2\sqrt{\left(1-\gamma_{\bq}/\m{J}\right)}
%\left(1-\xi\gamma_{\bq}\right)}
 \ee
where 
%$\xi= \left(\frac{\beta-\gamma}{\beta+\gamma}\right)^2$ measure the anisotropy while
 we have assumed for concreteness a hypercubic lattice in $d=z/2$ dimensions, 
\be 
\gamma_{\mathbf{q}} = \frac{1}{d}\sum_{a=1}^d\,\cos q_a \in [-1,1],\label{gamma_q}
\ee
with $q_a$ the components of the wavevector $\mathbf{q}$. 

The spectrum as expected is gapped on both sides of the transition
\be
\omega_{\bq}\simeq \omega^{\pm}_{gap}+\alpha_{\pm}\vert\bq\vert^2
\ee
 with a gap vanishing in a power-law fashion at the transition, 
 \be 
\omega^{\pm}_{gap}=c_{\pm}\vert J-J_c\vert^{1/2}
\ee 
Right at the phase boundary $J=J_c$ the spectrum becomes gapless with a linear dispersion $\omega_{\bq}=c\,\vert\bq\vert$ as expected for an Ising quantum critical point, whose dynamical critical exponent is $z_{dyn}=1$.

\subsubsection{Response to a Weak Driving and Photon Spectral Function}

Here we want to use the effective model to discuss the response of the Rabi-Hubbard model to a weak local drive and to compute the associated response function, i.e. the photon spectral function. Let us imagine driving a single resonator at position $\bR$ with a weak tone $\eta(t)$, i.e. to couple the system to the perturbation $V_{drive}(t)=\eta(t)\,x_{\bR}$  and measure the field at the same port $x_{\bR}(t)$. Using linear response theory we obtain
\be
\langle\, x_{\bR}(t)\rangle = \int\,dt'\chi_{loc}(t-t')\,\eta(t')
\ee
where the response function $\chi_{loc}(t)$ is the local retarded photon Green's function
\be
\chi_{loc}(t>0)=
%-i\langle\,[x_{\bR}(t),x_{\bR}(0)]\rangle=
-i\sum_{\bq}\, \langle[x_{\bq}(t),x_{-\bq}(0)]\rangle
\ee
It is therefore interesting to compute this response function in the frequency domain and discuss the signatures of the Rabi quantum phase transition. This can be done using the effective spin model we have previously discussed. We start considering the imaginary time correlator 
\be
\chi(\bq,\tau)=\langle\,T_{\tau}\,x_{\bq}(\tau)\,x_{-\bq}(0)\rangle
\ee
and then perform an analytic continuation.

 It is easy to see that within our effective model we have
$\chi(\bq,\tau)\sim \langle\,T_{\tau}\Sigma^{x}_{\bq}(\tau)\,\Sigma^{x}_{-\bq}(0)\rangle$
where the average and the imaginary time evolution is done with respect to the spin Hamiltonian~(\ref{eqn:Heff}). The above correlation can be evaluated using our harmonic theory of fluctuations. In particular, after performing the rotation~(\ref{eqn:rotation}) and bosonizing the spin operator as in~(\ref{eqn:bosonx}-\ref{eqn:bosonz}) we reduce the calculation to the evaluation of single particle correlators of harmonic oscillators. The result of this lengthy but straightforward calculation reads, in Matsubara frequencies $i\Omega_n=2\pi\,i\,n\,T$
\be\label{eqn:chi_mats}
%\chi(\bq,i\Omega)\sim\cos^2\theta\,C_{xx}(\bq,i\Omega)+\sin^2\theta\,C_{zz}(\bq,i\Omega) 
\chi(\bq,i\Omega)\sim\cos^2\theta\,A(\bq,i\Omega)+\sin^2\theta\,B(\bq,i\Omega) 
\ee
Here we have introduced the correlators
\be\label{eqn:Czz}
B(\bq,i\Omega)=K_{\bq}\left(\frac{1}{i\Omega+\omega_{\bq}}-\frac{1}{i\Omega-\omega_{\bq}}\right) 
\ee
and
\begin{widetext}
\be\label{eqn:Cxx}
A(\bq,i\Omega)=\langle\,x\rangle_0^2\,\delta(i\Omega)
+\left[\frac{K_{\bq}^2}{4}+\frac{1}{4\,K_{\bq}^2}-\frac{1}{2}\right]\,\left(n_{B}(\omega_{\bq})-n_B(-\omega_{\bq})\right)\left(\frac{1}{i\Omega+2\omega_{\bq}}-\frac{1}{i\Omega-2\omega_{\bq}}\right) 
\ee
\end{widetext}
with the bosonic occupation factor $n_B(\omega)=(e^{\beta\,\omega}-1)^{-1}$,
while $\langle\,x\rangle_0$ is the average $Z_2$ order parameter (including classical and quantum fluctuation contributions)
\be
\langle\,x\rangle_0=2\cos\theta\,\sum_{\bq}\,
\left[1-\frac{1}{2}\left(K_{\bq}+\frac{1}{K_{\bq}}\right)\right]
\ee
In addition we have introduced the kernel $K_{\bq}$ which reads for $J<J_c$
\be 
K_{\bq}=\frac{1}{\sqrt{1-\m{J}\gamma_{\bq}}}
\ee
while in the ordered phase $J>J_c$
\be
K_{\bq}=\frac{1}{\sqrt{1-\gamma_{\bq}/\m{J}}} 
\ee
With the above results at hand we now perform an analytic continuation onto the real axis $i\Omega\rightarrow\Omega+i\eta$ and sum over momenta to obtain the local retarded photon
Green's function, whose imaginary part defines our spectral function 
\be
\mbox{Im}\,\chi_{loc}(\Omega)= \sum_{\bq}\,\mbox{Im}\,\chi(\bq,i\Omega\rightarrow\Omega+i\eta)
\ee
Let's discuss now these results at zero temperature $T=0$ across the Rabi Hubbard phase diagram. 
\begin{figure}[t]
\begin{center}
\epsfig{figure=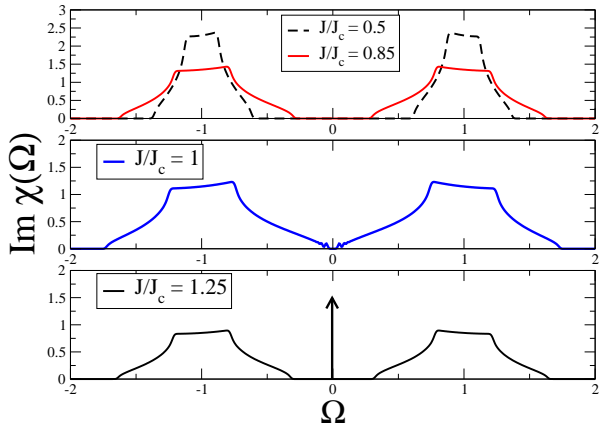,scale=0.325}
\caption{Imaginary part of the photon spectral function $Im\chi_{loc}(\Omega)$ for different values of the hopping strength across the critical point. In the insulating disordered phase $J<J_c$ the spectrum is gapped (top panel) and turns gapless at the quantum critical point $J=J_c$ with the spectrum vanishing as a power law at low frequency. Finally in the ordered phase $J>J_c$ a gap opens and a sharp coherent peak emerges at $\Omega=0$ signalling the breaking of the $Z_2$ symmetry.} 
\label{fig:photon_spect}
\end{center}
\end{figure}
In the disorderd phase, $J<J_c$, we have $\cos\theta=0$ and the photon spectrum contains two quasi-particle peaks at $\Omega=\pm\omega_{\bq}$, see Eq.~(\ref{eqn:Czz}). These give rise in the local spectrum to two broad high-energy features at frequencies $\Omega\in [\omega_-,\omega_+]$ and $\Omega\in [-\omega_+,-\omega_-]$ with edges  $\omega_{\mp}=\Delta\,\sqrt{1\mp\,\m{J}}$. Those two Rabi-Hubbard bands are separated by a gap $\omega_{gap}=2\omega_{-}$. For $J<J_c$ we obtain
\be
 \mbox{Im}\,\chi_{loc}(\Omega) = \theta\left(\Omega^2-\omega^2_-\right)\,\theta(\omega^2_+ -\Omega^2)\,\frac{\Delta}{\vert\Omega\vert}\,\rho_0\left(\frac{\Delta^2-\Omega^2}{\Delta^2-\omega_-^2}\right)
\ee
where $\rho_0(\omega)$ is the tight-binding Density of States (DoS) of the underlying photonic lattice.
The top panel of figure~\ref{fig:photon_spect} shows the photon spectral function in the Rabi insulating disordered phase. Right at the quantum critical point, $J=J_c$, the gap vanishes and the spectrum of excitations is gapless, $\omega_{\bq}=c\vert \bq\vert$. The photon spectral function can be obtained in closed form at this point and reads
\be
\mbox{Im}\,\chi_{loc}(\Omega) =\theta(\omega^2_+ -\Omega^2)\,\frac{\Delta}{\vert\Omega\vert}\,\rho_0\left(\frac{\Delta^2-\Omega^2}{\Delta^2}\right)
\ee
 Finally as the phase boundary is crossed and one enters the ordered phase at $J>J_c$ a new feature in the spectrum appears, related to the onset of a broken symmetry. In particular a sharp coherent delta peak at $\Omega=0$ appears in the middle of a gap (see Eq.~\ref{eqn:Cxx}) whose strength $Z$ is controlled by the order parameter.  In addition other quasiparticle peaks appear in the spectral function at higher frequencies $\Omega=\pm\,2\omega_{\bq}$ which result into an additional contribution to the high energy background of the local spectrum. As a result the ordered phase photon spectral function features two contributions, a coherent one at zero frequency and an incoherent term at higher frequencies
\be
\mbox{Im}\,\chi_{loc}(\Omega) = Z\,\delta(\Omega)+\mbox{Im}\,\chi^{incoh}_{loc}(\Omega) 
\ee
where the strength of the delta peak is given by
$$
Z= \left[\m{J}^2-1\right]\,\langle\,x\rangle_0^2
$$
The incoherent contribution is gapped and starts at the threshold $\omega_{-}\sim\sqrt{\m{J}-1}$. We plot the photon spectral function at the critical point as well as in the ordered phase in the lower panels of figure~\ref{fig:photon_spect}.

\section{Discussions and Conclusions}\label{sect:conclusions}

In this paper, we discuss the phase transition of light in a lattice of cavity QED systems. The basic building block of the lattice model we analyze here is the fundamental light-matter interaction Hamiltonian, the Rabi model. This model was previously discussed extensively by dropping the counter-rotating terms and introducing a chemical potential, the origin of which has been left unjustified -- this is the Jaynes-Cummings-Hubbard model. We argue in the present manuscript that the neglected counterclockwise terms are relevant operators, changing completely the nature of the phase transition of these interacting light-matter systems. In particular, we show that the counterclockwise terms act as an effective chemical potential stabilizing, out of vacuum, finite-density quantum phases of correlated photons and material excitations.\\

This changes completely the nature of quantum criticality of lattice CQED systems from those based on the Jaynes- Cummings model. While the latter is within the universality class of the Bose-Hubbard model, the Rabi-Hubbard model is a $Z_2$ parity-breaking quantum phase transition, where the two level systems polarize to generate a ferroelectrically ordered state and the photon coherence acquires a nonvanishing expectation value due to hopping. This quantum phase transition shares some aspects of the Dicke super-radiant critical point, with the addition of non trivial dynamical and quantum fluctuations that arise from the multi-mode nature of the photonic spectrum. From the point of view of its critical behaviour, the Rabi-Hubbard phase transition belongs to the universality class of the Quantum Ising model as we have shown here by introducing a low energy description, valid in the regime of large light-matter coupling, in terms of an emergent $Z_2$ low energy doublet.

Interesting research directions that we leave for future work concern, for example, the role of dissipation and photon losses on the Rabi-Hubbard phase diagram, or the possible signatures of the underlying quantum criticality in terms of experimentally measurable quantities. In this respect, owing to the simple discrete nature of the $Z_2$ symmetry of the problem, a possible direction could involve the non-adiabatic crossing of the phase transition and the observation of the resulting dynamics of defects formation, similar to what has been recently discussed in the context of 
 trapped ion systems~\cite{KibbleZurek_Ions1,KibbleZurek_Ions2}.
Finally, a particularly appealing research direction is the design of driven CQED architectures for realization of effective Hamiltonians with counter-rotating terms.

\textit{Acknowledgements -} We thank C. Ciuti, A. Clerk, S. Girvin, A. Houck, D. Huse, K. Le Hur, J. Keeling, 
F. Marquardt, D. Sadri and D. Underwood for stimulating discussions. This work was supported by the National Science Foundation through the Princeton Center for Complex Materials under Grant No. DMR-0819860 and through the NSF CAREER Grant No. DMR-1151810 and by the Swiss NSF through grant no. PP00P2-123519/1.

%\appendix

%\bibliography{../../../../mybiblio}

\end{document}